\documentclass[conference]{IEEEtran}

\IEEEoverridecommandlockouts
\usepackage{cite}
\usepackage{amsmath,amssymb,amsfonts}
\usepackage{graphicx}
\usepackage{textcomp}
\usepackage{xcolor,soul,framed} 
\usepackage{mdwtab}
\usepackage{eqparbox}
\usepackage{array}
\usepackage[caption=false,font=normalsize,labelfont=sf,textfont=sf]{subfig}
\usepackage{textcomp}
\usepackage{stfloats}
\usepackage{url}
\usepackage{verbatim}
\usepackage{cite}
\usepackage{multirow}
\usepackage{indentfirst}
\usepackage{booktabs}
\usepackage{breqn}
\usepackage{bm}
\usepackage{makecell}

\def\BibTeX{{\rm B\kern-.05em{\sc i\kern-.025em b}\kern-.08em
T\kern-.1667em\lower.7ex\hbox{E}\kern-.125emX}}
\usepackage[ruled]{algorithm2e}
\usepackage{algpseudocode}
\newtheorem{theorem}{\textbf{Theorem}}

\newtheorem{proposition}[theorem]{\textbf{Proposition}}
\newtheorem{remark}[theorem]{\textbf{Remark}}

\usepackage{url}
\usepackage{hyperref} 
\usepackage{xcolor}

\hypersetup{
    colorlinks=true, 
    linkcolor=blue, 
    filecolor=black, 
    urlcolor=blue, 
    citecolor=blue, 
    linkbordercolor={1 1 1}, 
    pdfborder={0 0 0} 
}
\graphicspath{ {./plots/} }
\begin{document}

 \title{A Lagrangian-Informed Long-Term Dispatch Policy for Coupled Hydropower and Photovoltaic Systems
\thanks{This work was partly supported by the National Science Foundation Graduate Research Fellowship under Grant No. DGE-2036197 and partly by the Department of Energy under Grant No. DE-EE0011385.}
}
\author{\IEEEauthorblockN{Eliza Cohn, Ning Qi, Upmanu Lall, Bolun Xu} 
\IEEEauthorblockA{\textit{Earth and Environmental Engineering} \\
\textit{Columbia University}\\
New York, USA\\
\{ec3766, nq2176, ula2, bx2177\}@columbia.edu}
\vspace{-0.5cm}
}

\maketitle

\begin{abstract}
This paper presents a long-term dispatch framework for coupled hydropower and floating photovoltaic systems. We introduce a temporal decomposition algorithm based on partial Lagrangian relaxation to address long-term water contract constraints. We derive a real-time, non-anticipatory dispatch policy based on water contract pricing.
Our framework is evaluated with a case study using real-world hydrology and power system data from Lake Mead and Lake Powell, on the Colorado River,  demonstrating competitive performance against commercial solvers for both linearized and nonlinear reservoir models. We conduct a sensitivity analysis on transmission capacity, electricity price and uncertainty scenarios, showing that the operational performance is significantly impacted by the transmission capacity and electricity prices while remaining relatively robust under uncertainty scenarios.


\end{abstract}

\section{Introduction}

Reservoir-based hydropower is an economical and clean form of electricity. It is a flexible form of a generation capable of rapidly deploying energy from stored water while incurring no additional fuel costs and producing negligible emissions~\cite{Singh2017-ay}. Recent deployments of floating solar photovoltaic (FPV) systems in hydropower reservoirs enhance the utilization of local power infrastructure, improve generation efficiency, and support a reliable and clean energy supply~\cite{Woolway2024-na,Rauf2020-ww}. This synergy enables innovative water release policies that enhance resilience and effectively integrate water management, solar generation, and grid requirements~\cite{Peng2015-oi}.

Operating hydropower with FPV must balance hourly power generation with water supply and flood control from the reservoir, requiring adherence to monthly water release targets and resulting in a multi-time-period optimization problem over a one-month horizon. However, solving this problem is challenging due to model nonlinearity~\cite{dogan2021hybrid},  long-term inter-temporal water release constraints~\cite{Li2014-eh}, etc. Regarding hydropower modeling, existing studies typically employ linear models~\cite{huang2023cascade}, piecewise linear models~\cite{tong2013milp}, or machine learning models~\cite{zhang2025long}. However, for large-scale multiple hydropower systems, these models become inefficient due to infeasibility or computational efficiency issues. On the other hand, existing studies primarily focus on short-term operations of coupled hydropower and FPV within a 24-hour horizon, addressing uncertainties using risk-averse methods with pre-known knowledge of uncertainties~\cite{qi2023chance,apostolopoulou2018robust} or rolling-horizon approaches with updated short-term forecasts~\cite{hu2022soft,ye2023real}. However, these methods become intractable for long-term operations, as long-term forecasts are often non-anticipated or unreliable~\cite{zhu2024short}, and decomposition methods struggle to manage long-term inter-temporal constraints. Moreover, the existing heuristic policies for long-term operations~\cite{shang2017improved} may lead to suboptimal solutions and reliability issues. The updated long-term reference, coordinated with online convex optimization, is proposed for long-term operations of hydrogen in~\cite{qi2025long}, but it does not account for long-term constraints. Hence, it is significant to develop a more sophisticated hydropower model and long-term decision-making policy for the coupled hydropower and FPV.


Motivated by this background, this paper proposes a Lagrangian-informed long-term dispatch framework with nonlinear and nonconvex hydropower model for coupled hydropower and FPV systems. Specially, our contributions of this paper are as follows:

\begin{itemize}
    \item \textbf{Methodology:} We develop a sequential non-anticipatory decision-making method for multi-time-period dispatch of coupled hydropower and FPV. The proposed method leverages partial Lagrangian relaxation to decompose the original problem into single-horizon dispatch and employs a bisection algorithm to determine the optimal shadow price of the long-term water release constraint. This enables efficient and reliable real-time energy management of coupled systems over a monthly horizon.
    \item \textbf{Modeling:} We incorporate a nonlinear and nonconvex hydraulic head function, fitted from real-world data, into the long-term dispatch framework. We demonstrate that this incorporation is both feasible and computationally efficient with the proposed decomposition method.
    \item \textbf{Numerical Analysis:} We validate the proposed framework through a real-world case study on the Hoover and Glen Canyon Dams, modeled in aggregate with a hypothetical FPV installation~\cite{Bureau-of-Reclamation2004-if}. We show that the proposed method achieves near-optimal performance with perfect foresight. We then conduct a sensitivity analysis on transmission
capacities and uncertainty scenarios which provides valuable insights on applications and further extension.
\end{itemize}


The remainder of the paper is organized as follows. Section~\ref{formulation} presents the problem formulation. Section~\ref{method} introduced the partial Lagrangian
relaxation method. Section~\ref{case} describes the numerical studies to demonstrate the effectiveness of the proposed method. Finally, we conclude this paper in Section~\ref{conclusion}. 

\section{Problem Formulation}\label{formulation}

We consider a coupled generation system combining hydropower and FPV. The coupled generation system operates under the same operator and transmission line. Hydropower manages reservoir water release for energy generation and other purposes, while FPV controls solar power based on available solar radiation. While the formulation proposed here is deterministic, we will show in Section~\ref{case} that it serves as a baseline for developing a non-anticipatory dispatch policy. 

\subsection{Formulation}
We formulate a multi-period optimization problem in~\eqref{constraints} to maximize total generation revenue over the time period set \( \mathcal{T} = \{1\text{,} 2\text{,} \dots\text{,} T\} \) based on price of electricity $\lambda_t$ (\$/MWh). All units are normalized by the time step duration which is one hour in our case study.
\begin{subequations}\label{constraints}
\begin{align}
&\max_{h_t\text{,}s_t\text{,}u_t\text{,}V_t} \sum_{t \in \mathcal{T}} {\lambda_t(h_t + s_t)} \label{objfunction}\\ 
\text{s.t. }& \sum_{t \in \mathcal{T}}u_t = U : \theta\label{watercontract} \\
& V_t - V_{t-1} = I_t - u_t\text{, }t \in \mathcal{T} \label{massbal} \\
& 0 \leq h_t \leq \eta g \rho \phi(V_{t-1}) u_t/3.6e^9\text{, }t \in \mathcal{T} \label{powerfunc} \\
& \underline{u} \leq u_t \leq \overline{u}\text{, }t \in \mathcal{T} \label{flowrate} \\
& -\underline{R} \leq u_t - u_{t-1} \leq \overline{R}\text{, }t \in \mathcal{T} \label{ramprate} \\
& 0 \leq s_t \leq \alpha_t S\text{, }t \in \mathcal{T} \label{solarcap} \\
& 0 \leq s_t + h_t \leq P\text{, }t \in \mathcal{T} \label{feedercap}
\end{align}
\end{subequations}
\noindent where decision variables include  $s_t$ (MWh) denotes the dispatched energy from FPV at time period \(t\); $h_t$ (MWh) denotes the dispatched energy from hydropower at time period \(t\); $u_t$ ($\mathrm{m^3}$) denotes the hourly water release at time period \(t\); $V_t$ ($\mathrm{m^3}$) denotes the reservoir volume at time period \(t\). 


\eqref{watercontract} models the monthly water contract $U$ ($\mathrm{m^3}$) and $\theta$ is the associated shadow price variable. \eqref{massbal} tracks the water balance in the reservoir system based on the inflow $I_t$ ($\mathrm{m^3}$). The hydropower generation from water release is limited by~\eqref{powerfunc}, in which $\eta$ denotes the hydropower generation efficiency; $\phi(\cdot)$ (m) denotes the hydraulic head function; $g$ denotes the gravitational constant; $\rho$ ($\mathrm{kg/ m^3}$) denotes the density of water. A conversion from joule to MWh (1MWh=$3.6e^9$Joule) ensures unit consistency on both sides of~\eqref{powerfunc}. 
\eqref{flowrate} models the upper and lower water release bounds $\underline{u}$ and $\overline{u}$, and~\eqref{ramprate} models the water flow ramp rates $\underline{R}$ and $\overline{R}$, all in $\mathrm{m^3}$. \eqref{solarcap} limits the solar power based on the available solar radiation with the capacity factor $\alpha_t$ and the installed solar capacity $S$ (MWh). \eqref{feedercap} limits the total generation from hydro and FPV with the transmission capacity $P$ (MWh).

\subsection{Fitting the Nonlinear Hydraulic Head Function}
\begin{figure}[!ht]
\setlength{\abovecaptionskip}{-0.1cm}  
    \setlength{\belowcaptionskip}{-0.1cm} 
    \centering
    \includegraphics[width=0.85\columnwidth]{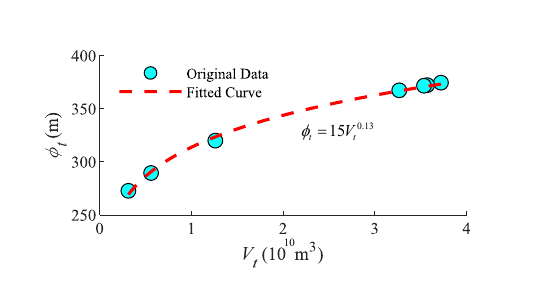}
    \caption{Visualization of fitting the nonlinear hydraulic head function.}
    \label{fig:theta_k}
\end{figure}

We incorporate the nonlinear hydraulic head function for hydropower, assuming a concave relationship between reservoir volume and height. 
\begin{equation}\label{head}
    \phi(V_t) = aV_t^b 
\end{equation}
\noindent where parameters $a$ and $b$ are obtained by fitting real-world data\footnote{https://www.usbr.gov/lc/region/g4000/LM
AreaCapacityTables2009.pdf},
as illustrated in Fig.~\ref{fig:theta_k}. The $R^2$ value from the line of best fit is 0.99. Preserving nonlinearity in the hydraulic head mirrors the real-world hydrology dynamics as opposed to simplifications introduced by traditional approximations such as a piece-wise linear functions or  convex counterparts~\cite{Santosuosso2024-xg}. Since all operations are conducted within the range of the fitted data, interpolating points within the model remains reliable.

\section{Methodology}\label{method}

We propose a partial Lagrangian relaxation approach to decompose the inter-temporal constraint of the water contract delivery. This allows the optimization problem to be solved independently at each time step.

\subsection{Partial Lagrangian Relaxation}
The formulation in \eqref{constraints} has only one long-term time coupling constraint \eqref{watercontract}. All other constraints only consider a single or two neighboring time steps, while the objective is also separable. Hence, we utilize the Lagrangian relaxation of the long-term water contract to generate a two-level formulation:
\begin{subequations}\label{reformulation}
\begin{align}
    &\min_{\theta} \Big(U-\sum_{t \in \mathcal{T}}u^*_t \Big)^2 \text{ where } \label{lag1}\\
    & u^*_t \in \arg\max_{h_t\text{,} s_t\text{,} u_t\text{,}V_{t}} \Big\{\lambda_t(h_t+s_t) - \theta u_t \text{ s.t. \eqref{massbal} - \eqref{feedercap}}\Big\} \label{lag2}
\end{align}
\end{subequations}
where $u^*_t$ represents the optimized water release from the time-decoupled subproblem. The subproblem \eqref{lag2} is a non-anticipatory which only optimizes the generation and water release during time step $t$ based on the input water contract price $\theta$, while the master objective \eqref{lag1} ensures the water contract is fulfilled by the lower-level non-anticipatory reservoir operations by find the optimal $\theta$. Hence, the master problem is a water contract pricing problem, while the lower-level subproblems simulate realistic reservoir operations. Note that the minimized master problem will be zero representing the original water release contract \eqref{watercontract}. Since the water contract is established on a monthly basis, $\theta$ remains constant throughout the duration of each month. 

This decomposed formulation provides two advantages compared to \eqref{constraints}. First, it simulates non-anticipatory reservoir operation, allowing the water contract to be priced in a realistic manner. Each subproblem only includes decision variables at the current step and considers past and current information without looking into the future, as \eqref{massbal}-\eqref{feedercap} only includes the current time step and all previous time steps. Second, it significantly reduces the computation complexity in solving the nonlinear hydraulic head function in \eqref{powerfunc}. Note that solving \eqref{constraints} over a typical water contract duration, such as one month, at hourly resolution can be extremely computation-challenging with an arbitrary hydraulic head function $\phi$, but in the decomposed structure we only need to solve one step at a time and the optimization can be efficiently solved.

\subsection{Bi-section Solution Method of the Master Problem}

We leverage the fact that the master problem \eqref{lag1} has only a single unknown $\theta$. Based on this, we propose a bisection algorithm to find the optimal solution. The bisection algorithm relies on the following proposition:

\begin{proposition}
    The sum of the water releases, $\sum_{t \in \mathcal{T}} u^*_t$, decreases monotonically with respect to $\theta$ if the hydraulic head function $\phi$ is constant.
\end{proposition}

The proof of this proposition is as follows. First, we observe from \eqref{lag2} that $\theta$ represents the price associated with water release. The upper bound of hydropower generation $h_t$ is proportional to the water release $u_t$, as modeled by the hydraulic head function in \eqref{powerfunc}. Therefore, as $\theta$ increases, the effective price of water release also increases, leading to a reduction in $u_t$ over each time step $t$.

There are two complicating factors in this proposition, the ramp constraint \eqref{ramprate} and the hydraulic head function \eqref{powerfunc}. Since the subproblem \eqref{lag2} simulates non-anticipatory reservoir operation, we perform the analysis in sequential order. For the ramp constraint, we start from the first period $t=1$, and it is trivial to see what the proposition holds. Then, in the next time period, a higher value of $\theta$ not only further penalizes the water release but also reduces the water release upper bound due to the ramp constraint \eqref{ramprate} because the earlier water release must also be lower. Hence, the proposition holds for the \eqref{ramprate}. While for the hydraulic head function, we have to assume it to be a constant, hence \eqref{powerfunc} be a linear constraint for the proposition to hold rigorously, hence we finished the proof.

\begin{remark}
    For nonlinear hydraulic head function, we show that Proposition~1 may not necessarily hold by substituting \eqref{powerfunc} into the objective of \eqref{lag2} assuming it is binding. The objective then becomes
    \begin{align}
        \max_{h_t\text{,} s_t\text{,} u_t\text{,}V_{t-1}} \lambda_ts_t + (\lambda_t\eta g \rho \phi(V_t)/3.6e^9  - \theta) u_t
    \end{align}
    Note that a lower water release in prior periods due to higher $\theta$ will lead to a higher $V_t$ hence $\phi(V_t)$ will have a higher value. Hence, $\lambda_t\eta g \rho \phi(V_t)/3.6e^9  - \theta$ may not change monotonically with $\theta$. Hence, we cannot conclude that Proposition~1 will definitely hold. However, we will show later in the simulation that using realistic data, the proposed algorithm provides similar results to benchmark IPOPT solvers.
\end{remark}

Building on this proposition, we introduce an auxiliary variable, $\sigma(\theta)$, to represent the total water release as a function of the  $\theta$. We formally define this as:
\begin{equation}
    \sigma(\theta) = \sum\nolimits_{t \in \mathcal{T}} u^*_t, \label{watercontract2}
\end{equation}
\noindent where $u^*_t$ is the solution obtained from \eqref{lag2}.

We can now develop the algorithm by picking an arbitrary $\theta$ and simulate the hourly reservoir operation as stated in \eqref{lag2}. Assume $\theta^*$ is the minimizer to \eqref{lag1} we have:
\begin{align}
&\text{If } \sigma(\theta) > U \rightarrow \theta < \theta^* \text{, }  \text{If } \sigma(\theta) < U \rightarrow \theta > \theta^*\text{, } \label{cond} \\
&\text{If } \sigma(\theta) = U \rightarrow \theta = \theta^* \nonumber
\end{align}

\subsection{Analytical Solution to the Subproblem}
We now propose an analytical algorithm for solving the decomposed subproblem \eqref{lag2} for fast and efficient computation without the need for commercial solvers especially regarding the nonlinear hydraulic head function. 
\begin{proposition}
    The solution to \eqref{lag2} with a given $\theta$ is as follows if $\lambda_t \geq 0$:
    \begin{subequations}\label{prop3}
\begin{align}
    s^*_t &= \min\{\alpha_tS,P\} \label{prop3:1} \\
    u^*_t &= \max\{u_{t-1}-\underline{R}, \underline{u}, \min\{\overline{u}, u_{t-1}+\overline{R}, \hat{u_t}\}\} \label{prop3:2}\\
    V^*_t &= V_{t-1} + I_t - u^*_t \label{prop3:3}\\
    h^*_t &= \min\{P-s^*_t, \eta g \rho \phi(V^*_t) u^*_t\} \label{prop3:4}
\end{align}
where 
\begin{align}
\hat{u_t} &= \begin{cases}
    \dfrac{(P-s^*_t)\cdot3.6e^9}{\eta g \rho \phi(V_{t-1})} & \text{if $\hat{\theta} > \theta$} \label{prop3:5}\\
    0 & \text{else}
\end{cases}\\
    \hat{\theta} &= \lambda_t\eta g \rho \phi(V_{t-1})/3.6e^9 \label{prop3:6}
\end{align}
\end{subequations}
\end{proposition}

The proof of this proposition is as follows. We start by assuming $\lambda_t \geq 0$ then it is trivial to see the shadow price of \eqref{watercontract} $\theta$ should also be a non-negative value and the right-hand side of \eqref{powerfunc} must always be binding due to the positive value of releasing water and \eqref{powerfunc} is the only constraint coupling $h_t$ and $u_t$. Since $s_t$ has no marginal cost, then $s_t$ should always be upper bounded by the lesser of \eqref{solarcap} and \eqref{feedercap} before considering $h_t$ as in \eqref{prop3:1}. \eqref{flowrate} and \eqref{ramprate} are both box constraints in the decomposed subproblem - note $u_{t-1}$ is treated as a parameter. Then \eqref{prop3:5} calculates $u_t$ without these two box constraints based on the coefficient of $u_t$ by subsisting \eqref{powerfunc} into the objective to replace $h_t$ as in \eqref{watercontract2}, in which $u_t$ would reach the transmission capacity minus $s_t$ if the weight is positive. Finally, \eqref{prop3:2} enforces the box constraints back over $u_t$ and we update $V_t$ and $h_t$ accordingly. 

\subsection{Full Algorithm}
We implement a bi-section search tree algorithm to iteratively determine the optimal value of $\theta$ and the corresponding dispatch solutions. A confident search range is to choose $L = 0$, $R = 1$ and set $\epsilon = 10^{-6}$.

\begin{algorithm}[htbp]\label{bst}
\hspace*{\fill}
\caption{Bi-section Search Tree Algorithm}
\SetAlgoLined
\SetAlgoNoEnd 
\KwIn{Initial search range $[L\text{,} R]$ and accuracy $\epsilon$.}
\KwOut{Optimal price of water $\theta^*$.}
\SetKwBlock{StepOne}{Search Algorithm:}{}
\StepOne{
    \While{$R-L > \epsilon$}{
        Set $\theta = (L + R)/2$ \\
        Update subproblem solutions as in~\eqref{prop3} sequentially from $t=1$ to $T$\\
        Calculate $\sigma(\theta)$ as in \eqref{watercontract2} \\
        \If{$\sigma(\theta) > U$}{
        $R \leftarrow \theta$ (increase the penalty)
        }
        \Else{
        $L \leftarrow \theta$ (decrease the penalty)
        }
        \textbf{end}
    }
    Return $\theta$ as the optimal price of water. 
}
\end{algorithm}

\section{Numerical Studies}\label{case}

\subsection{Set-up}
We demonstrate the effectiveness of the proposed method through real-world case studies on the coupled hydropower and FPV systems at Lake Powell and Lake Mead, USA. Draining and overtopping of the reservoir are not considered in the model, as the large reservoir capacity sufficiently governs the release dynamics. 
The hourly electricity price data is from CAISO,
the hourly available solar radiation is from NREL,
and the daily water inflow data from the US Bureau of Reclamation\footnote{CAISO: http://oasis.caiso.com; NREL: https://nsrdb.nrel.gov; \\ USBR: https://www.usbr.gov/uc/rm/crsp/gc/}
over 2022-2023 is collected and available online\footnote{[Code Available]: https://github.com/ecohn44/fpv-hydro-dispatch}. The system configurations and parameters are shown in Table \ref{tab:parameters}, which are derived from~\cite{Bureau-of-Reclamation-Upper-Colorado-Region2016-mb} and \cite{Spencer2019-nn}. The optimization is coded in Julia and the programming environment is an Apple M3 Core @ 4.06 GHz with RAM 16 GB.


\begin{table}[!ht]
\renewcommand{\arraystretch}{1.2}
\caption{System Parameters}
\label{tab:parameters}
\centering
  \setlength{\tabcolsep}{5mm}{\begin{tabular}{c c c c}
\toprule
Parameter & Value & Parameter & Value \\
\midrule
$\overline{u}$ & 707.9 \(\text{m}^3\) & $\underline{u}$ & 141.6 \(\text{m}^3\) \\
$\overline{R}$ & 113.3 \(\text{m}^3\) & $\underline{R}$ & 70.4 \(\text{m}^3\) \\
$P$ & 1300 MW & $S$ & 1000 MW \\
$\eta$ & 0.775 & $g$ & 9.8 $\text{m}\text{/s}^2$ \\
$\rho$ & 1000 $\text{kg}/\text{m}^3$ & $T$ & 730 $\text{h}$ \\
\bottomrule
\end{tabular}}
\end{table}

\subsection{Comparative Results}
We compare the results with different methodologies:
\begin{enumerate}
    \item M1: Multi-period optimization as~\eqref{constraints}.
    \item M2: Proposed decomposed optimization as~\eqref{reformulation}.
\end{enumerate}


The original multi-period formulation (M1) is not tractable with a nonlinear hydraulic head for the full duration of the water contract. Gurobi solves only the linearized version and IPOPT has a sharp degradation in performance for any horizon larger than two weeks. The advantages of our proposed algorithm (M2) is that the nonlinear hydraulic is  feasible over any finite time periods. Because the full month cannot be solved for M1 and M2, they are benchmarked via IPOPT over a one week period in January of 2022. The cumulative revenue under M1 and M2 is 10.013 M\$ and 10.012 M\$, the total water release is 1.696 x$10^8$ m$^3$ and 1.695 x$10^8$ m$^3$, and the cpu time is 5039.1 ms and 17.6 ms respectively. We can conclude that M2 is nearly optimal because (i) the revenue of M1 is only 0.1\% larger and (ii) M1 only releases 0.06\% more water than M2. Furthermore, M2 sees significant advantages in its improved computational efficiency. 

The methodologies are next bench-marked with a linearized hydraulic head via Gurobi over the same one week period. The cumulative revenue under M1 and M2 is 10.29 M\$ and 10.10 M\$, the total water release is 1.696 x$10^8$ m$^3$ and 1.694 x$10^8$ m$^3$, and the cpu time is 19.8 ms and 17.4 ms respectively. The conclusions of near-optimality are the same as above. 


\subsection{Sensitivity Analysis}
\begin{figure}[!ht]
\setlength{\abovecaptionskip}{-0.1cm}  
    \setlength{\belowcaptionskip}{-0.1cm} 
    \centering
\includegraphics[width=0.9\columnwidth]{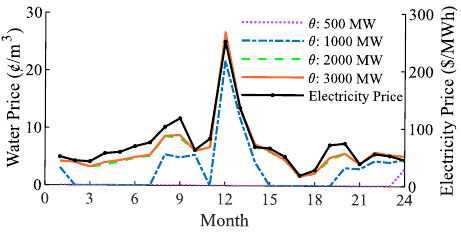}
    \caption{Shadow price of water ($\theta$) and average monthly marginal price of electricity ($\lambda_t$) for varying transmission capacities (P) over 2022-2023. }
    \label{fig:WP_LMP}
\end{figure}

In Figure \ref{fig:WP_LMP}, the shadow price of water solved for in Alg. \ref{bst} is overlaid with the average price of electricity for each month. The correlation coefficient between the two prices ranges from 0.86 and 0.98 under a transmission capacity of 1 to 3 GW, demonstrating how the price of electricity drives the dispatch of both FPV generation and water release.

\begin{figure}[!ht]
\setlength{\abovecaptionskip}{-0.1cm}  
    \setlength{\belowcaptionskip}{-0.1cm} 
    \centering
\includegraphics[width=0.9\columnwidth]{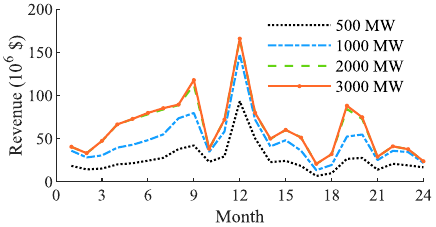}
    \caption{Revenue for varying transmission capacity (P) from 2022 to 2023.}
    \label{fig:rev_FC}
\end{figure}

Figure \ref{fig:WP_LMP} shows that the shadow price of water increases proportionally with transmission capacity. This indicates that increased capacity for generation increases the value of water released. However, there is a limit in this potential value since the shadow price of water for 2 GW and 3 GW is nearly the same. 
For a small transmission capacity rating such as 0.5 GW, the system can rely solely on FPV generation to drive profits and the water used for generation becomes less valuable. Figure \ref{fig:rev_FC} shows that revenue also increases proportionally with transmission capacity, but there is a limit in total earnings as the transmission capacity nears 3 GW. 


Under the deterministic formulation, a knowledge of the next period average price of electricity means it is possible to predict the shadow price and control release under near-optimal operating guidelines. 
This is advantageous because the average monthly price is far easier to predict than the higher resolution forecasts of inflow, solar radiation, or hourly price of electricity used in standard hydropower operations. 

\subsection{Sensitivity Study on Uncertainty Incorporation}\label{uncertainty}

The practical application of the proposed dispatch framework must incorporate uncertainties, including solar radiation, water inflow, and electricity price. In this case, the reservoir operator can use predicted stochastic scenarios to solve the water contract shadow price using the developed method, while the proposed control policy is non-anticipatory and, therefore, applicable to a stochastic operation setting. 

\begin{figure}[!ht]
\setlength{\abovecaptionskip}{-0.1cm}  
    \setlength{\belowcaptionskip}{-0.1cm} 
    \centering
\includegraphics[width=0.9\columnwidth]{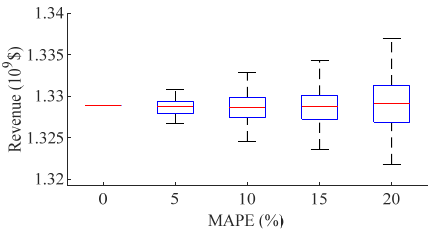}
    \caption{Total revenue distribution over varying MAPE from 2022 to 2023.}
    \label{fig:MC_Sims}
\end{figure}

To evaluate the feasibility and performance robustness of the proposed method for incorporating uncertainties, we conduct Monte Carlo simulations by introducing autoregressive (AR(1)) noise to electricity price, solar radiation, and water inflow while ensuring the long-term water contract constraint is still enforced.
The prediction uncertainty is quantified by the mean absolute percentage error (MAPE). Utilizing the shadow price of water derived from the deterministic formulation, we perform Monte Carlo simulations across MAPE levels of 0\%, 5\%, 10\%, 15\%, and 20\%. Figure~\ref{fig:MC_Sims} records the total revenue accumulated over a two-year period from these simulations. The results demonstrate that the pre-determined shadow price remains a reliable factor in guiding operational decisions when incorporating uncertainties. The case with a MAPE level of 0\% can represent the performance of the online optimization methods, while cases with non-zero MAPE levels reflect the performance of rolling-horizon approaches. Notably, at the highest 20\% MAPE error setting, the change of revenue is less than 1\% in all scenarios, demonstrating our method can provide robust performance against operational uncertainties. 

\section{Discussion and Conclusion}\label{conclusion}



This paper proposes a Lagrangian-informed long-term dispatch framework for coupled hydropower and FPV systems. The proposed framework leverages partial-Lagrangian relaxation and the shadow price of long-term water release constraint, as well as incorporates the nonlinear and nonconvex hydropower model. We validate the effectiveness of the proposed method through a real-world case study, demonstrating that it achieves near-optimal performance compared to the perfect foresight method, with an optimality gap of less than 0.1\%. Sensitivity analysis further confirms that the method ensures reliable, robust, and efficient real-time energy management of coupled systems under varying conditions. Future work will extend our method to incorporate the stochastic optimization approach and multiple coupled hydropower and FPV systems.




\bibliographystyle{IEEEtran}
\bibliography{paperpile}

\end{document}